\documentclass[showpacs,preprintnumbers,amsmath,amssymb]{revtex4}
\usepackage[dvips]{graphicx}
\usepackage{epsfig}
\usepackage{soul}

\begin{document}

\title{Qubit-Qutrit ($2 \otimes 3$) quantum systems: An investigation of some quantum correlations under collective dephasing}
\author{Mazhar Ali}
\affiliation{Department of Electrical Engineering, Faculty of Engineering, Islamic University Madinah, 107 Madinah, Saudi Arabia}

\begin{abstract}
We revisit qubit-qutrit quantum systems under collective dephasing and answer some of the questions which have not been asked and addressed so far in the 
literature. In particular, we examine the possibilities of non-trivial phenomena of {\it time-invariant} entanglement and {\it freezing} dynamics of 
entanglement for this dimension of Hilbert space. Interestingly, we find that for qubit-qutrit systems both of these peculiar features coexist, that is,  
we observe not only time-invariant entanglement for certain quantum states but we find also find evidence that many quantum states freeze their 
entanglement after decaying for some time. To our knowledge, the existance of both these phenomena for 
one dimension of Hilbert space is not found so far. All previous studies suggest that if there is freezing dynamics of entanglement, then there is 
no time-invariant entanglement and vice versa. In addition, we study local quantum uncertainity and other correlations for certain families of states 
and discuss the interesting dynamics. Our study is an extension of similar studies for qubit-qubit systems, qubit-qutrit, and multipartite quantum systems. 
\end{abstract}


\maketitle

\section{Introduction}
\label{Sec:intro}

Quantum correlations have their role in potential applications in quantum information theory. This includes remote state preparation \cite{Dakic-Nat-2012}, 
entanglement distribution \cite{Streltsov-PRL108-2012, Chuan-PRL109-2012}, transmission of correlations \cite{Streltsov-PRL111-2013}, and 
quantum metrology \cite{Modi-PRX1-2011} to name a few. This utilization of quantum correlations is already enough motivation to study, characterize 
and quantify them. There are classical correlations which have no quantum in it. Quantum correlations are difficult to characterize and quantify and there 
are several different techniques to capture them. 
Entanglement, quantum discord and local quantum uncertainity are kind of quantum correlations. Even there are different measures to compute entanglement 
and quantum discord. Nevertheless, these correlations have attracted lot of interest and considerable 
efforts have been devoted to develop a theory of these correlations \cite{Horodecki-RMP-2009, gtreview, Chiara-RPP2018}. 
The advancement in experimental setups during last couple of decades, enabled us to work for realistic realizations of quantum devices utilizing 
quantum correlations. 
Due to unavoidable interactions of delicate quantum systems with their environment, it is essential to simulate the effects of noisy environments 
on quantum correlations. Such investigations are already an active area of research \cite{Aolita-review} and several authors have studied decoherence 
effects on quantum correlations for both bipartite and multipartite systems
\cite{lifetime,Aolita-PRL100-2008,bipartitedec,Band-PRA72-2005,lowerbounds,Lastra-PRA75-2007, Guehne-PRA78-2008,Lopez-PRL101-2008, 
Ali-work, Weinstein-PRA85-2012,Ali-JPB-2014, Ali-2015, Ali-2016}. 

There are several types of experimental setups to test the ideas of quantum information. One of the technological advanced setup is to trap the 
ions/atoms and perform quantum computations by logic gates, measurements etc. In these experiments, the typical noise is caused by intensity 
fluctuations of electromagnetic fields which leads to collective dephasing process. 
This process degrades quantum correlations and there are already many investigations of the effects of collective dephasing on 
entanglement for bipartite and multipartite quantum systems 
\cite{Yu-CD-2002, AJ-JMO-2007,Li-EPJD-2007, Song-PRA80-2009, Ali-PRA81-2010, Karpat-PLA375-2011, Liu-arXiv, Carnio-PRL-2015, Carnio-NJP-2016, 
Ali-IJQI-2017, Ali-EPJD-2017}.  
It has been reported in these studies that collective dephasing process offers not only the expected exponential decay of entanglement but also the 
abrupt end of entanglement (sudden death of entanglement). In addition to these two dynamical behavior, some of the recent studies demonstrated that 
there are two other types of non-trivial dynamics of entanglement present/observed under collective dephasing. 
First, there is so called {\it time-invariant} entanglement \cite{Karpat-PLA375-2011, Liu-arXiv, Ali-EPJD-2017}. 
Time-invariant entanglement does not necessarily mean that the quantum states live in decoherence free subspaces (DFS). In fact the quantum states 
may change at every instance whereas their entanglement remain constant throughout the dynamical process. This feature was first observed for 
qubit-qutrit systems \cite{Karpat-PLA375-2011} and then later on observed for qubit-qubit systems as well \cite{Liu-arXiv}. 
Recently, we have investigated time-invariant phenomenon for genuine entanglement of three and four qubits and explicitly observed this 
phenomenon \cite{Ali-EPJD-2017}. 
The second non-trivial feature of entanglement decay is called {\it freezing} dynamics of entanglement 
\cite{Carnio-PRL-2015, Carnio-NJP-2016, Ali-IJQI-2017}. It was shown that a specific two qubits state may first decay upto some numerical value 
before suddenly stop decaying and maintain this stationary entanglement \cite{Carnio-PRL-2015, Carnio-NJP-2016}. 
Recently, we have explored freezing dynamics for various genuine multipartite specific states of three and four qubits, including random 
states and found evidence for it \cite{Ali-IJQI-2017}.  
More recently, we have explored the possibility of either time-invariant entanglement or freezing dynamics for qutrit-qutrit ($3 \otimes 3$) 
systems \cite{Ali-MPLA-2019}. We found no evidence for time-invariant entanglement, however we observed the exclusive evidence for 
freezing dynamics of entanglement \cite{Ali-MPLA-2019}. We have noticed that in all previous studies on time-invariant entanglement 
and freezing dynamics for a given Hilbert space, there is either time-invariant entanglement or freezing dynamics behavior. We have not found 
so far these two features occuring together for one dimension of Hilbert space. Interestingly, for qubit-qutrit systems we find both these 
features present. As we show below, there are certain states which exibit either time-invariant entanglement or sudden death of entanglement but 
never freezing dynamics. On the other hand, some other quantum states exhibit either freezing dynamics or sudden death but never time-invariant 
dynamics. However, we get both peculiar features for Hilbert space of dimension $6$. 

The two other quantum correlations which we study in this work are quantum discord and local quantum uncertainity. Quantum discord may be defined as the 
difference between quantum mutual information and classical correlations \cite{Ollivier-PRL88-2001, Vedral-et-al, Luo-PRA77-2008}. 
Quantum discord may be nonzero even for separable states and have applications in quantum information. Due to complicated minimization process, 
the computation of discord is not an easy task and analytical results are known only for some restricted families of states. 
For $2 \otimes d$ quantum systems, analytical results for quantum discord are known for a specfic family of states \cite{Ali-JPA43-2010} and 
the general procedure to calculate discord is also worked out \cite{Rau-2017}. The dynamics of quantum discord under decoherence has been 
studied \cite{Werlang-work} and is found to be more robust than quantum entanglement. In this work, we also study dynamics of quantum discord and 
classical correlation under collective dephasing for a specfic family of states. 
The other quantum correlation which we study in this work is recently proposed, known as local quantum uncertainity \cite{Girolami-PRL110-2013}. 
This measure is based on idea of skew information and it is discord type correlation \cite{Modi-RMP84-2012}. Recently, the effects of decoherence 
on discord-like measures including local quantum uncertainity has been studied \cite{Bera-RPP81-2018, Karpat-CJP96-2018, Slaoui-work}.  
Here in this work, we study local quantum uncertainity for several families of quantum states under collective dephasing. 
We find that in situations where entanglement exhibits time-invariant feature, local quantum uncertainity first keep on increasing to a 
specfic value and then exhibit freezing dynamics after long time. In instances, where there is entanglement sudden death, local quantum 
uncertainity first decays, then increase and finally tend to freeze in the long time. On the other hand, in situations where entanglement 
exhibits freezing dynamics, local quantum uncertainity first decays very slowly to a value and then decays abruptly and finally tend to exhibit 
freezing dynamics as well. Finally, we examine the random pure states and calculate their entanglement at infinity. We find that more than half random 
states main their entanglement at infinity and hence all other correlations as well under collective dephasing.  

This paper is organized as follows. In section \ref{Sec:Model}, we briefly discuss our model of interest and obtain the most general solution for 
an arbitrary initial density matrix. In section \ref{Sec: QCs}, we review the idea of entanglement for qubit-qutrit systems and describe the method 
to compute negativity for an arbitrary initial quantum state. We also briefly examine the concept of quantum discord and how to compute it for an arbitrary 
bipartite state. We also briefly review local quantum uncertainity and how to compute it for any state for $2 \otimes d$ quantum systems. 
In section \ref{Sec: Res}, we provide our main results for various initial states. Finally, we conclude our work in section \ref{Sec: Cc}.

\section{Collective dephasing for qubit-qutrit systems} 
\label{Sec:Model}

Our physical model consists of a qubit and a qutrit (one two-level atom and one three-level atom for an example) $A$ and $B$ that are coupled to a noisy 
environment, collectively. The qutrit as an atom, can be realized with well known "V"-type energy level configuration in which the transition among 
excited levels is forbidden. This means that first excited state will decay to ground level only and similarly the second excited level will also decay to 
ground level. The atoms are sufficiently far apart and they do not interact with each other, so that we can treat them as independent. The 
collective dephasing refers to coupling of atoms to the same noisy environment, which can be stochastic magnetic fields $B(t)$.
There are at least two approaches to write a Hamiltonian for such physical situations. First, the Hamiltonian could be time independent, like in 
case of a qubit $H = \hbar \omega/2 \, \sigma_z$ with $\omega$ as energy splitting between excited states of atom. One can write a unitary propagator  
$U(t) = \exp(- i H t/\hbar)$. As there are fluctuations in magnetic field strength, the integration over it will induce a probability distribution
$p(w)$ of characteristic energy splitting. The time evolution of atom can be written as an integral over $p(\omega)$ and unitary evolution, i.e., 
$\rho(t) = \int p(\omega) U(t) \rho(0) U(t)^\dagger \, d\omega$. The form of $p(\omega)$ will determine the nature of noise. Another approach, 
which we have taken in this work and most of the work in literature is to take the Hamiltonian as time dependent and embed the 
fluctuations of magnetic field in stochastic function $B(t)$, which already includes the information about characteristic function and 
so that the ensemble average over it introduce the decay parameter $\Gamma$. Both approaches are equivalent and generates the same dynamics. 
However, we point out, to our knowledge the present work and recent works are restricted to a very specific orientation of magnetic field and 
the theory of a general description of magnetic fields in any arbitrary directions is still not worked out.
The Hamiltonian of the quantum system (with $\hbar = 1$) can be written as \cite{Karpat-PLA375-2011} 
\begin{eqnarray}
H(t) = - \frac{\mu}{2} \, \big[ \, B(t) (\sigma_z^A + \sigma_z^B) \,  \big] \, , \label{Eq:Ham} 
\end{eqnarray}
where $\mu$ is gyromagnetic ratio and $\sigma_z^A$ is standard Pauli matrix for qubit and $\sigma_z^B$ is the dephasing operator for qutrit $B$. 
The stochastic magnetic fields refer to statistically independent classical Markov processes satisfying the conditions
\begin{eqnarray} 
\langle B(t) \, B(t')\rangle &=& \frac{\Gamma}{\mu^2} \, \delta(t-t') \,, \nonumber \\ 
\langle B(t)\rangle &=& 0 \, ,
\end{eqnarray}
with $\langle \cdots \rangle$ as ensemble time average and $\Gamma$ denote the phase-damping rate for collective dephasing.

Let $|2\rangle$, $|1\rangle$, and $|0\rangle$ be the first excited state, second excited, and ground state of the qutrit, respectively. 
We choose the computational basis $ \{ \, |0,0\rangle$, $|0,1\rangle$, $|0,2\rangle \, |1,0\rangle$, $|1,1\rangle$, $|1,2\rangle \, \}$, 
where we have dropped the subscripts $A$ and $B$ with the understanding that first basis represents qubit $A$ and second qutrit $B$. Also the notation 
$|0 \rangle \otimes |0\rangle = |0 \, 0 \rangle$ has been adopted for simplicity. 
The time-dependent density matrix for the system is obtained by taking ensemble average over the noisy field,
i.\,e., $\rho(t) = \langle\rho_{st}(t)\rangle$, where $\rho_{st}(t) = U(t) \rho(0) U^\dagger(t)$ and
$U(t) = \exp[-\mathrm{i} \int_0^t \, dt' \, H(t')]$. The most general solution of $\rho(t)$ under the assumption that the system is 
not initially correlated with environment is given as
\begin{widetext}
\begin{eqnarray}
\rho(t) = \left(
\begin{array}{llllll}
\rho_{11} & \xi \, \rho_{12} & \xi^4 \, \rho_{13} & \xi^4 \, \rho_{14} & \xi^9 \, \rho_{15} & \xi^{16} \,\rho_{16} \\
\xi \, \rho_{21} & \rho_{22} & \xi \, \rho_{23} & \xi \, \rho_{24} & \xi^4 \, \rho_{25} & \xi^9 \, \rho_{26} \\
\xi^4 \, \rho_{31} & \xi \, \rho_{32} & \rho_{33} & \rho_{34} & \xi \, \rho_{35} & \xi^4 \, \rho_{36} \\
\xi^4 \, \rho_{41} & \xi \, \rho_{42} & \rho_{43} & \rho_{44} & \xi \, \rho_{45} & \xi^4 \, \rho_{46} \\
\xi^9 \, \rho_{51} & \xi^4 \, \rho_{52} & \xi \, \rho_{53} & \xi \, \rho_{54} & \rho_{55} & \xi \, \rho_{56} \\
\xi^{16} \, \rho_{61} & \xi^9 \, \rho_{62} & \xi^4 \, \rho_{63} & \xi^4 \, \rho_{64} & \xi \, \rho_{65} & \rho_{66}
\end{array}
\right) \,,\label{Eq:MF}
\end{eqnarray}
\end{widetext}
where $\xi = e^{- \Gamma t/8}$. We note that decoherence free subspaces (DFS) \cite{Yu-CD-2002} do appear in this system as a common characteristic 
of collective dephasing. Another interesting property of the dynamics is the fact that all initially zero matrix elements remain zero.

\section{Entanglement, quantum discord and local quantum uncertainity for $2 \otimes 3$ quantum systems}
\label{Sec: QCs}

In this section, we briefly review the correlations, which we study in this work for qubit-qutrit systems. In 
subsection \ref{SS:ent}, we briefly review entanglement and a computable measures of entanglement. In subsection \ref{SS:dis}, we review the 
quantum discord and how to compute it for any bipartite quantum state. In subsection \ref{SS:lq}, we discuss local quantum uncertainity and how to 
compute it for a given state in $2 \otimes d$ quantum systems.  

\subsection{Quantum entanglement}
\label{SS:ent}

The question of quantum entanglement for qubit-qubit $(2 \otimes 2)$ quantum systems and qubit-qutrit $(2 \otimes 3)$ quantum systems has been solved. 
It is well known that for bipartite quantum systems, if the partial transpose with respect of any one of the subsystem has at least one negative 
eigenvalue then the quantum state is entangled or NPT \cite{Peres-PRL-1996}. Whereas if the partial transposed matrix has all positive eigenvalues (PPT), 
then entanglement/separability depends upon the dimension of Hilbert space. The PPT states for $2 \otimes 2$ and $2 \otimes 3$ are separable 
(not entangled), whereas for larger dimensions of Hilbert space, there may exist PPT-entangled states 
(also called bound entangled states) \cite{Horodecki-RMP-2009}. 
Hence, for a given density matrix of qubit-qutrit system, one can easily find the eigenvalues of partially transposed matrix (partial transpose can 
be taken with respect to any subsystem). It is not hard to look for possible negative eigenvalues. The sum of absolute values of all possible negative 
eigenvalues is defined as a legitimate measure of quantum entanglement, namely {\it negativity} \cite{Vidal-PRA65-2002}. Hence, negativity is defined as 
\begin{equation}
 N (\rho) = 2 \, \bigg( \, \sum_i \, | \eta_i| \, \bigg)\, ,
\end{equation}
where $\eta_i$ are possible negative eigenvalues and multiplication with $2$ is for normalization so that for maximally entangled states, this measure 
should have numerical value of $1$. For specific quantum states, this definition is sufficient to compute and study the dynamics of negativity. For random 
states, it is more easy to use {\it entanglement monotone}, which is based on PPT-mixtures idea \cite{Bastian-PRL106-2011} and very easy to compute 
numerical value of entanglement for any density matrix. The description of semi-definite programming (SDP) and genuine negativity is described in details in 
Ref.\cite{Bastian-PRL106-2011}. We denote this measure by $E(\rho)$ in this paper. For bipartite systems, this monotone is equivalent to {\it negativity}. 

\subsection{Quantum Discord}
\label{SS:dis}

Quantum discord is one of the measure of quantum correlations which are captured using von Neumann entropy. This measure has been intensively 
investigated in previous 18 years in various contexts and many studies focused on the quantification of this measure for various dimensions of Hilbert space. 
The literature on this measure is so extensive that it is not possible to cite each of them, so we only provide fundamental references. We discuss the 
main ideas very briefly to compute quantum discord for a given bipartite quantum state. 
Any bipartite state may have both quantum and classical correlations, which are jointly captured by quantum mutual information. 
In particular, if $\rho^{AB}$ denotes the density operator of a composite bipartite system $AB$,
and $\rho^A$ ($\rho^B$) the density operator of part $A$ ($B$), respectively, then the quantum mutual information is defined as \cite{Groisman-PRA72-2005}
\begin{eqnarray}
\mathcal{I} (\rho^{AB}) = S (\rho^A) + S (\rho^B) - S(\rho^{AB})\, , \label{Eq:QMI}
\end{eqnarray}
where $S(\rho) = - \mathrm{tr} \, ( \rho \, \log_2 \rho )$ is the von Neumann entropy. We take all logarithms base $2$ in this work. 
Quantum mutual information may be written as a sum of classical correlation $\mathcal{C}(\rho^{AB})$ and quantum discord $\mathcal{Q} (\rho^{AB})$, 
that is, $\mathcal{I} (\rho^{AB}) = \mathcal{C} (\rho^{AB}) + \mathcal{Q} (\rho^{AB})$ \cite{Ollivier-PRL88-2001, Vedral-et-al, Luo-PRA77-2008}. 
Quantum discord can be positive in separable mixed states (that is, with no entanglement).

Quantum discord can be quantified \cite{Ollivier-PRL88-2001} via von Neumann type measurements which consist of 
one-dimensional projectors that sum to the identity operator. Let the projection operators $\{ A_k\}$ describe a von Neumann measurement for 
subsystem $A$ only, then the conditional density operator $\rho_k$ associated with the measurement result $k$ is 
\begin{eqnarray}
\rho_k = \frac{1}{p_k} (A_k \otimes \mathbb{I}_B) \, \rho \, (A_k \otimes \mathbb{I}_B) \,,
\end{eqnarray}
where the probability $p_k$ equals $\mathrm{tr} [(A_k \otimes \mathbb{I}_B) \, \rho \, (A_k \otimes \mathbb{I}_B)]$.
The quantum conditional entropy with respect to this measurement is given by \cite{Luo-PRA77-2008} 
\begin{eqnarray}
S (\rho | \{A_k\}) := \sum_k p_k \, S(\rho_k) \, ,
\label{Eq:QCE}
\end{eqnarray}
and the associated quantum mutual information of this measurement is defined as
\begin{eqnarray}
\mathcal{I} (\rho|\{A_k\}) := S (\rho^B) - S(\rho|\{A_k\}) \, . \label{Eq:QMIM} 
\end{eqnarray}
A measure of the resulting classical correlations is provided \cite{Ollivier-PRL88-2001, Vedral-et-al, Luo-PRA77-2008} by 
\begin{eqnarray}
\mathcal{C}(\rho) := \sup_{\{A_k\}} \, \mathcal{I} (\rho|\{A_k\}) \, . \label{Eq:CC} 
\end{eqnarray}
The obstacle to computing quantum discord lies in this complicated maximization procedure for calculating the classical
correlation because the maximization is to be done over all possible von Neumann measurements of $A$. Once $\mathcal{C}(\rho)$ is in hand, quantum discord 
is simply obtained by subtracting it from the quantum mutual information,
\begin{eqnarray}
\mathcal{Q}(\rho) := \mathcal{I}(\rho) - \mathcal{C}(\rho) \, .
\end{eqnarray}
This maximization process is not easy in general and analytical results for quantum discard are only known for very specific quantum states. In this work, 
we have been only able to calculate it for only one family of quantum states for $2 \otimes 3$ quantum system. 

\subsection{Local quantum uncertainity}
\label{SS:lq}

First of all we briefly review the concept of local quantum uncertainity (LQ). This is a measure of quantum correlations which has been defined for 
$2 \otimes d$ quantum systems \cite{Girolami-PRL110-2013}. It is a quantum discord type measure and we will see in the results below that 
for certain quantum states, quantum discord and local quantum uncertainity captures precisely same correlations and are equal to each other, 
whereas for some other states, they are different measures. It is defined as the minimum skew information which is obtained via local measurement on 
qubit part only. This measure has the advantage that there is no need for complicated minimization over parameters related with measurement operations. 
This measure is defined as 
\begin{equation}
LQ(\rho) \equiv \, \min_{K_A} \, \mathcal{I} (\rho , K_A \otimes \mathbb{I}_B) \,, 
\end{equation}
where $K_A$ is some local observable on subsystem $A$, and $\mathcal{I}$ is the skew information of the density operator $\rho$, defined as 
\begin{equation}
 \mathcal{I} (\rho , K_A \otimes \mathbb{I}_B) \, = \,- \frac{1}{2} \, \rm{Tr} ( \, [ \, \sqrt{\rho}, \, K_A \otimes \mathbb{I}_B ]^2 \, ) \,.
\end{equation}
It has been shown \cite{Girolami-PRL110-2013} that for $2 \otimes d$ quantum systems, the compact formula for local quantum uncertainity is given as
\begin{equation}
LQ(\rho) = 1 - \rm{max} \, \{ \lambda_1 \,, \lambda_2 \, , \lambda_3 \, \}\,, 
\end{equation}
where $\lambda_i$ are the eigenvalues of $3 \times 3$ matrix $\mathcal{M}$, whose matrix elements are calculated by relationship
\begin{equation}
 m_{ij} \equiv \rm{Tr} \, \{ \, \sqrt{\rho} \, (\sigma_i \otimes \mathbb{I}_B) \, \sqrt{\rho} \, (\sigma_j \otimes \mathbb{I}_B) \, \}\,, 
\end{equation}
where $i,j = 1,2,3$ and $\sigma_i$ are the standard Pauli matrices.

\section{Main results}
\label{Sec: Res}

In this section, we will present our main results for various families of quantum states. 

\subsubsection{Two parameter class of states}

The class of quantum states with two real parameters $\alpha$ and $\gamma$ in a $2 \otimes d$ quantum system \cite{Chi-JPA36-2003} is given as 
\begin{eqnarray}
\rho_{\alpha, \gamma} =& \alpha \, \sum_{i = 0}^{1} \sum_{j = 2}^{d-1} \, | i\, j \rangle\langle i \, j| + \beta \, ( | \phi^+ \rangle\langle \phi^+| 
+ | \phi^- \rangle\langle \phi^-| + | \psi^+ \rangle\langle \psi^+| \, ) \nonumber \\&  + \gamma \, | \psi^- \rangle\langle \psi^-| \,,
\label{Eq:rhoag}
\end{eqnarray}
where $\{ \, |i \, j \rangle : \, i = 0, \, 1, j = 0, \, 1, \, \ldots \, , \, d-1 \, \}$ is an orthonormal basis for $2 \otimes d$ quantum system and
\begin{eqnarray}
| \, \phi^\pm \rangle &=& \frac{1}{\sqrt{2}} \, ( \, |0\, 0\rangle \pm |	1\,1 \rangle \, ) \\
| \, \psi^\pm \rangle &=& \frac{1}{\sqrt{2}} \, (\, |0\,1 \rangle \pm |1\,0 \rangle ) \, , 
\end{eqnarray}
and the parameter $\beta$ is dependent on $\alpha$ and $\gamma$ by the unit trace condition, 
\begin{eqnarray}
 2 \, ( d - 2) \alpha + 3 \, \beta  + \gamma = 1 \, .
\end{eqnarray}
From Eq.~(\ref{Eq:rhoag}) one can easily obtain the range of parameters as $ 0 \leq \alpha \leq 1/(2 (d-2))$ and $0 \leq \gamma \leq 1$. We note that 
the states of the form $\rho_{0, \gamma}$ are equivalent to Werner states \cite{Wer-PRA89} in a $2 \otimes 2$ quantum systems. Moreover, 
the states $\rho_{\alpha, \gamma}$ have the property that their PPT (positive partial transpose) region is always separable \cite{Chi-JPA36-2003}. 
It is also known that an arbitrary quantum state $\rho$ in $2 \otimes d$ can be transformed to $\rho_{\alpha, \gamma}$ with the help of 
local operations and classical communication (LOCC).

We have already calculated quantum discord, classical correlation and entanglement for this family in an earlier work \cite{Ali-JPA43-2010}. Here we simply 
extend the previous results for collective dephasing (an additional parameter $\Gamma t$). 
It turns out that classical correlations for this family of states does not depend on decay parameter and are constant in time. The expression for 
classical correlations is given as
\begin{eqnarray}
\mathcal{C}(\rho_{\alpha,\gamma}) = - (3 \, \beta + \gamma) \, \log(\frac{3 \, \beta + \gamma }{2}) + 2 \, \beta \, \log(2 \, \beta) 
+ (\beta + \gamma) \log(\beta + \gamma) \, . 
\label{Eq:cc23} 
\end{eqnarray}
The quantum discord is calculated using the standard procedure discussed in previous section and is given as
\begin{eqnarray}
\mathcal{Q}(\rho_{\alpha,\gamma})(t) =& 1 - 2 \, \alpha -2 \, \beta -(\beta + \gamma) \, \log(\beta + \gamma) + \frac{\beta + \gamma 
+ \xi \, (\beta-\gamma)}{2} \, \log\bigg(\frac{\beta + \gamma + \xi \, (\beta - \gamma)}{2}\bigg) 
\nonumber \\& + \frac{\beta + \gamma - \xi \, (\beta - \gamma)}{2} \, \log\bigg(\frac{\beta + \gamma - \xi \, (\beta - \gamma)}{2}\bigg) \,. 
\label{Eq:qd23}
\end{eqnarray}
We can see that as $t \to \infty$, $\xi \to 0$, and $\mathcal{Q}(\rho_{\alpha,\gamma})(\infty) = 1 - 2 \, \alpha - 3 \, \beta - \gamma = 0$ as expected.

The local quantum uncertainity for this family of state turns out to be 
\begin{eqnarray}
LQ(\rho_{\alpha,\gamma})(t) = 1 - 2 \, \alpha - 2 \, \beta - \bigg[\sqrt{\beta (1 + \xi) + \gamma (1 - \xi)} \, \sqrt{\beta \, (1 - \xi) 
+ \gamma \, (1 + \xi)} \bigg]\,.
\label{Eq:lqab1}
\end{eqnarray}
We note the similarity between local uncertainity Eq.(\ref{Eq:lqab1}) and quantum discord Eq.(\ref{Eq:qd23}). Indeed, it turns out that for $t = 0$, 
and for the initial states (i) $\alpha = \beta = 0$ and $\gamma = 1$, (ii) $\alpha = \gamma = 0$, and $\beta = 1/3$, (iii) $\gamma = 0$, and 
(iv) $ \beta = 0$, local quantum uncertainity and quantum discord turns out be exactly equal as can be checked easily. However, for more general cases with 
$\alpha, \beta, \gamma \neq 0$, and under collective dephasing, both measured are different as will shown below. 

The negativity for this family of states is straight forward to calculate and is given as
\begin{equation}
N (\rho_{\alpha, \gamma})(t) = \rm{max} \big[ \, 0 \, , \xi \, (\gamma - \beta) - 2 \, \beta \big] \,.
\end{equation}
It is easy to see that for $\beta = 0$, the states decay asymptotically and entanglement is lost only at infinity, whereas for $ \beta \neq 0$, 
negativity is lost at  
\begin{equation}
\Gamma t =  8 \, \log \frac{\gamma - \beta}{ 2 \, \beta} \, .
\label{Eq:Nabe}
\end{equation}

\begin{figure}[t!]
\scalebox{2.20}{\includegraphics[width=1.95in]{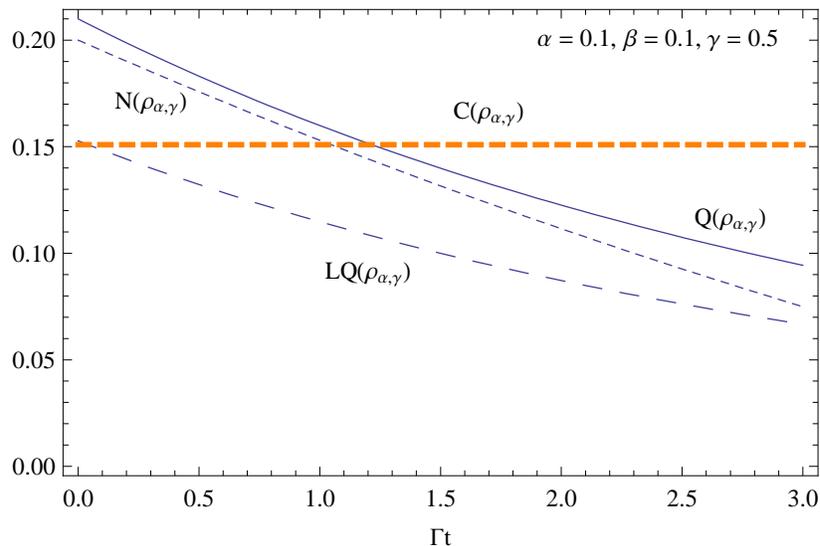}}
\caption{Entanglement (negativity) $N (\rho_{\alpha, \gamma})$, classical correlation $C (\rho_{\alpha, \gamma})$, quantum discord 
$\mathcal{Q}(\rho_{\alpha, \gamma})$, and local quantum uncertainity $LQ (\rho_{\alpha, \gamma})$ are plotted against 
parameter $\Gamma t$. It can be seen that all correlations maintain nonzero values for long time due to presence of decoherence free subspace.}
\label{Fig:1}
\end{figure}

We plot entanglement, discord, classical correlation, and local quantum uncertainity for state $\rho_{\alpha, \gamma}(t)$ in Figure~(\ref{Fig:1}). 
We have taken specific values of $\alpha = 0.1$, $\beta = 0.1$, and $\gamma = 0.5$. Quantum discord $ \mathcal{Q} (\rho_{\alpha, \gamma})(t)$ plotted 
as solid line decays slowly as well as negativity (dashed line) and local quantum uncertainity (big dashed line). Classical correlation 
(dashed orange line) is constant in time with fixed initial value. Negativity ends at $\Gamma t \approx 5.54$ (not shown in Figure~\ref{Fig:1}), whereas 
quantum discord becomes zero at infinity. Local quantum uncertainity and quantum discord become zero at the same time as expected.  

\begin{figure}[t!]
\scalebox{2.20}{\includegraphics[width=1.95in]{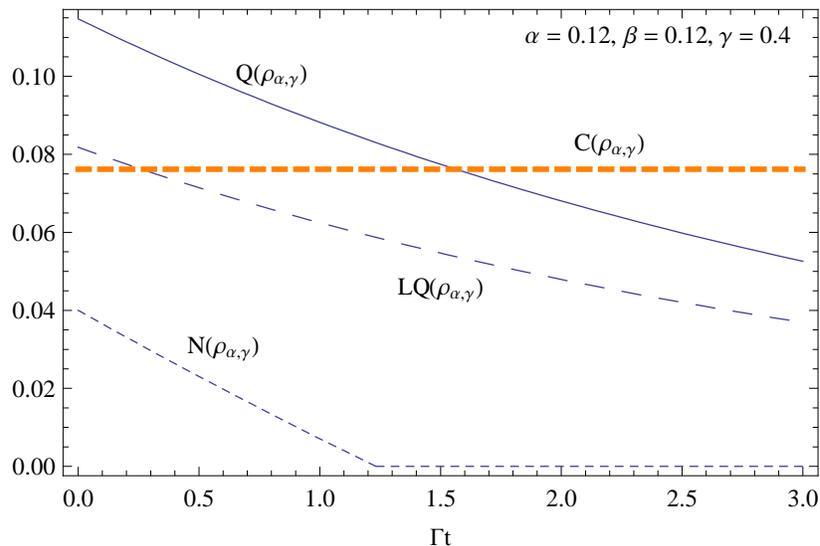}}
\caption{Entanglement (negativity) $N (\rho_{\alpha, \gamma})$, classical correlation $C (\rho_{\alpha, \gamma})$, quantum discord 
$\mathcal{Q}(\rho_{\alpha, \gamma})$, and local quantum uncertainity $LQ (\rho_{\alpha, \gamma})$ are plotted against 
parameter $\Gamma t$. It can be seen that all correlations except negativity maintain nonzero values for long time.}
\label{Fig:2}
\end{figure}

Let us take another set of initial values with $\alpha = 0.12$, $\beta = 0.12$, and $\gamma = 0.4$ for state $\rho_{\alpha, \gamma}(t)$. 
Figure~(\ref{Fig:2}) depicts entanglement (dashed line), classical correlation (thick dashed orange line), quantum discord (solid line) and 
local quantum uncertainity (big dashed line) for this set of values against decay parameter $\Gamma t$. As we have reduced the fraction of 
maximally entangled state ($\gamma$) and increased the noisy components $\alpha$ and $\beta$ slighly, nevertheless, the resulting dynamics is 
interesting and different than earlier case. The numerical values of all correlations are lower than the earlier case. This fact is understandable 
as we have reduced the fraction of $\gamma$, so maximally entangled state feeded almost all correlations in $\rho_{\alpha, \gamma}$. 
Another main difference is vanishing of entanglement at $\Gamma t \approx 1.233$ so called sudden death of entanglement. 
Classical correlation is constant as mentioned earlier. Quantum discord and local quantum uncertainity decaying slowly as expected and both becoming 
zero only at infinity.

\subsubsection{Search for freezing dynamics of entanglement}

It has already been shown explicitly \cite{Karpat-PLA375-2011} that certain qubit-qutrit entangled states exhibit time-invariant entanglement 
feature under collective dephasing. However, the question of freezing dynamics of entanglement has not been explored so far. Therefore, we look for such 
possibility encouraged by the existance of decohence free subspaces where entangled states can reside. Of course, alone the presence of such decoherence 
free spaces do not guarantee that either time-invariant entanglement or freezing dynamics must occur. In fact all previous studies suggest that for all 
other dimensions of Hilbert space studied so far, either time-invariant entanglement appear or freezing dynamics. To our knowledge, these both 
possibilities have never been observed for any single dimension of Hilbert space. Interestingly, as we will demonstrate that qubit-qutrit systems 
offer all kind of dynamical features of entanglement, that is, entanglement sudden death, asymptotic decay of entanglement, time-invariant entanglement, 
and freezing dynamics of entanglement under collective dephasing. 

Let us define a single parameter class of states, which are mixture of entangled states residing in decoherence free subspace and states which decay. 
The states are defined as
\begin{equation}
\rho_\alpha =  \alpha \, |\psi_3 \rangle\langle \psi_3| + (1 - \alpha) \, |\psi_2 \rangle\langle \psi_2| \,,    
\end{equation}
where $0 \leq \alpha \leq 1$, the maximally entangled state $|\psi_2 \rangle$ is defined as
\begin{equation}
|\psi_2 \rangle = \frac{1}{\sqrt{2}} \, (|0 \, 1\rangle + |1 \, 2 \rangle)\,, 
\end{equation}
and another maximally entangled state $|\psi_3 \rangle$ is defined as
\begin{equation}
|\psi_3 \rangle = \frac{1}{\sqrt{2}} \, (|0 \, 2\rangle + |1 \, 0 \rangle)\, . 
\end{equation}
In this mixture $|\psi_2 \rangle$ decays, whereas $|\psi_3 \rangle$ lives in decoherence free subspace. Therefore the time evolution of this state 
can be written as
\begin{equation}
\rho_\alpha (t) = \alpha \, \rho_3 + (1 - \alpha) \, \rho_2 (t) \, .    
\end{equation}
There are only two possible negative eigenvalues for the partial transpose of this state, namely 
\begin{eqnarray}
v_1(\alpha) = \frac{1}{4} \, \bigg[ (1-\alpha) - \sqrt{(1-\alpha)^2 + 4 \, \alpha^2} \bigg] \nonumber \\
v_2 (\alpha,\,\xi) = \frac{1}{4} \, \bigg[ \alpha - \sqrt{\alpha^2 + 4 \, \xi^{18} \, (1- \alpha)^2} \bigg] \,.
\end{eqnarray}
Negativity for these states can be written as
\begin{equation}
 N_\alpha = 2 \, \big[ \, {\rm max} ( 0, \, - v_1(\alpha) ) + {\rm max} (0, \,  - v_2(\alpha)) \, \big]\, .
\end{equation}
It is obvious that $v_1(\alpha)$ does not depend on decay parameter and this value is negative for any $\alpha > 0$. The other eigenvalue 
$v_2(\alpha, \xi)$ is also negative for any $\alpha > 0$ at the start ($\Gamma t = 0$), however, as decoherence is turned on, this value quickly becomes 
positive. So we can see very clearly that all states with $ 0 < \alpha < 1$ must exhibit freezing dynamics of entanglement. 

\begin{figure}[t!]
\scalebox{2.20}{\includegraphics[width=1.95in]{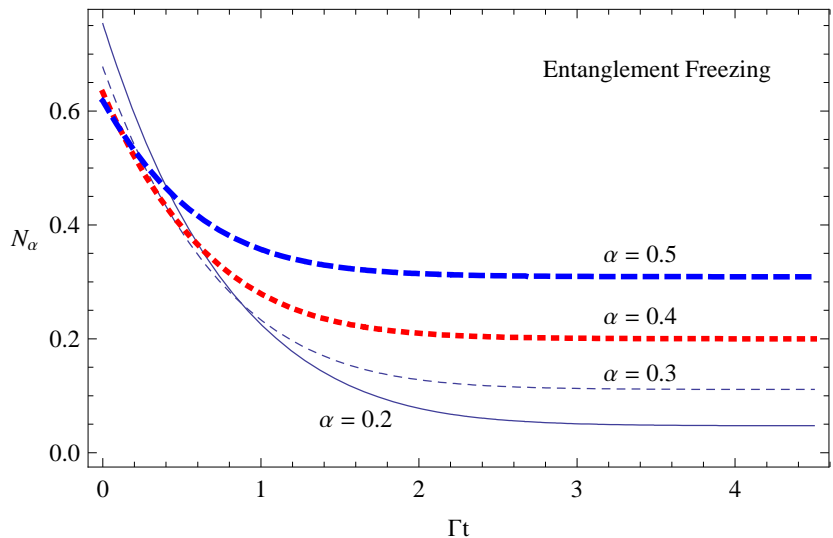}}
\caption{Negativity for an initial state $\rho_{\alpha}(t)$ is plotted against decay parameter $\Gamma t$ for various values of parameter $\alpha$. 
It can be seen that initial entanglement decays to a specific value (depending on $\alpha$) and then although quantum states keep changing with time 
but entanglement becomes stationary hence exhibiting so called freezing dynamics of entanglement. See text for explanations.}
\label{Fig:3}
\end{figure}
Figure (\ref{Fig:3}) shows negativity plotted against decay parameter $\Gamma t$ for various choices of parameter $\alpha$. It is clear that all initial 
amounts of entanglement determined by choice of $\alpha$ decay as evident from $v_2(\alpha)$ until it becomes zero and hence the residual entanglement in 
decoherence free subspace becomes dominant as dictated by $v_1(\alpha)$. Hence this family of states exhibit freezing dynamics of entanglement such that 
quantum states changes with time but its entanglement is locked in time (stationary). It is interesting to note that for qubit-qutrit systems, 
time-invariant entanglement and freezing dynamics exist. We have not found this coincidence in any other dimension of Hilbert space so far.   

It is straight forward to calculate local quantum uncertainity for $\rho_\alpha$ which is given as
\begin{equation}
LQ_\alpha = 1 - \frac{1}{2} \, \sqrt{\alpha (1-\alpha)} \,. 
\end{equation}
This value is symmetric about $\alpha = 0.5$ as expected because all correlations must be symmetric about this value. For time-evolved state, local 
quantum uncertainity is given as 
\begin{equation}
LQ_\alpha(t) = 1 - \lambda_\alpha(t) \,, 
\end{equation}
where $\lambda_\alpha(t) = \rm{max} [w_{11}(t), w_{33}(t)]$, and 
\begin{equation}
w_{11}(t) = w_{22}(t) = \frac{\sqrt{\alpha} \, \bigg[ \sqrt{(1-\alpha)(1 - \xi^9)} + \sqrt{(1 - \alpha)(1 + \xi^9)} \bigg]}{2 \sqrt{2}} \, , 
\end{equation}
and 
\begin{equation}
w_{33}(t) = (1-\alpha) \, \sqrt{1 - \xi^{18}} \, , 
\end{equation}
where $w_{ii}(t)$ are the eigenvalues of the symmetric $3 \times 3$ matrix.
\begin{figure}[t!]
\scalebox{2.20}{\includegraphics[width=1.95in]{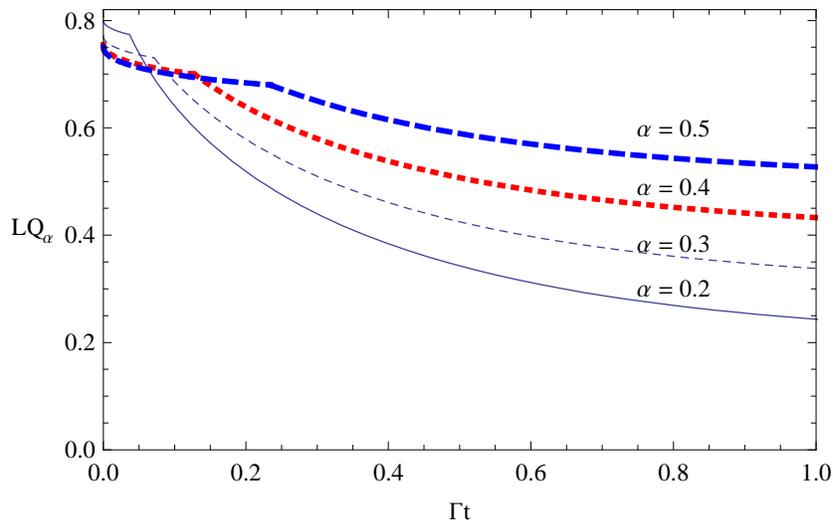}}
\caption{Local quantum uncertainity is plotted against parameter $\Gamma t)$ for different values of parameter $\alpha$. It can be 
seen that first $LQ_\alpha$ decay but then tend to be become stationary.}
\label{Fig:4}
\end{figure}
In Figure (\ref{Fig:4}) we plot the local quantum uncertainity against decay parameter $\Gamma t$ for same values of parameter $\alpha$ as in 
Figure~(\ref{Fig:3}). As can be seen that just like entanglement freezing, the local quantum uncertainity initially decays to some value and then 
also tend to freezing dynamics of local uncertainity. At $\Gamma t = \infty$, the stationary value of local quantum uncertainity is given as
\begin{equation}
LQ_\alpha (\infty) = 1 - \rm{max} \bigg[ (1-\alpha),\, \sqrt{\alpha (1-\alpha)/2} \bigg]\, ,  
\end{equation}
which is abvious a nonzero value. 

\subsubsection{A review on time-invariant entanglement for qubit-qutrit systems}

As we have noticed in all earlier reports of time-invariant entanglement, the quantum state exhibiting this interesting phenomenon must be a mixture of 
two entangled states and one of the state must reside in decoherence free subspace. However, we have seen above that if we mix state 
$|\psi_3 \rangle$ and $|\psi_2 \rangle$, we do not obsereve any time-invariant entanglement rather freezing dynamics of entanglement. So this suggests 
that we must look for some other entangled state to be mixed with $|\psi_3 \rangle$. One of such state is 
\begin{equation}
|\psi_1 \rangle = \frac{1}{\sqrt{2}} \, (|0 \, 0\rangle + |1 \, 2 \rangle)\, .
\end{equation}
Actually the first report of time-invariant entanglement for qubit-qutrit systems \cite{Karpat-PLA375-2011} took a state which was mixture of 
these two type of states. To generalize this observation for more general states, first let us consider the states,
\begin{equation}
\tilde{\rho}_\alpha = \alpha \, |\psi_1 \rangle\langle \psi_1| + \frac{1 - \alpha}{6} \, \mathbb{I}_6 \,, 
\end{equation}
where $\mathbb{I}_6$ is $6 \times 6$ identity matrix and $ 0 \leq \alpha \leq 1$. Such states are called isotropic states and they are NPT for 
$1/4 < \alpha \leq 1$, and hence entangled. To avoid confusion, we differentiate these state by taking tilde over $\rho_\alpha$. This could have been 
avoided by calling single parameter any other name than $\alpha$, however we preferred to keep it like that. 
We can now define two parameter family of states, which are mixture of isotropic states and $|\psi_3\rangle$, given as 
\begin{equation}
\rho_{\alpha,\beta} = \beta \, |\psi_3\rangle \langle \psi_3 | + (1 - \beta) \, \tilde{\rho}_\alpha \, ,    
\end{equation}
where $ 0 \leq \beta \leq 1$. Entanglement properties for this family of states are quite interesting. The partial transpose with respect to 
subsystem $A$ have maximum two possible negative eigenvalues and the rest of $4$ eigenvalues are definitely 
positive for the given range of parameters $\alpha$ and $\beta$. The $2$ possible negative eigenvalues are such that when one is positive, then 
other is negative and vice versa. They are never negative at the same time.  
The time evolution of these states can be written as
\begin{equation}
\rho_{\alpha,\beta}(t) = \beta \, |\psi_3\rangle \langle \psi_3 | + (1 - \beta) \, \tilde{\rho}_\alpha(t) \, .    
\end{equation}
Hence $\tilde{\rho}_\alpha(t)$ decays, whereas $|\psi_3\rangle$ remain dynamically invariant as it lives in DFS. Now there is an additional 
parameter $\Gamma t$ involved in the density matrix. The two possible negative eigenvalues of partially transposed matrix are given as 
\begin{eqnarray}
x_1 &=& \frac{1}{6} \, \big[ 1 + 2 \, \alpha (1-\beta) - 4 \, \beta \, \big] \nonumber \\ 
x_2 &=& \frac{1}{6} \, \big[ 1 + 2 \, \beta - \alpha \, (1-\beta)(1 + 3 \, \xi^{16} ) \, \big]\, .    
\label{Eq:NEVs}
\end{eqnarray}

As we have mentioned earlier that these two eigenvalues can not be negative at the same time. We also observe that one of the eigenvalue $x_1$ does not 
depend upon $\xi$, so if this eigenvalue is negative then as the other cannot be negative so this necessary means time-invariant entanglement. On 
the other hand if $x_1$ is positive then $x_2$ must be negative. However $x_2$ depends on $\xi$ and it not difficult to see that $x_2$ can become 
positive in a finite time, leading to finite time end of entanglement. As long as $\beta > 1/2$, $x_2$ is positive for all ranges of $\alpha$, hence 
we can get time-invariant entanglement, whereas for other values we would get sudden death of entanglement. Negativity for these state is given as
\begin{equation}
 N_{\alpha\,, \beta} = 2 \, \big[ \, {\rm max} \big( 0, \, - x_1(\alpha,\beta) \big) + {\rm max} \big(0, \,  - x_2(\alpha , \beta, \xi) \big) \, \big]\, .
\end{equation}

\begin{figure}[t!]
\scalebox{2.20}{\includegraphics[width=1.95in]{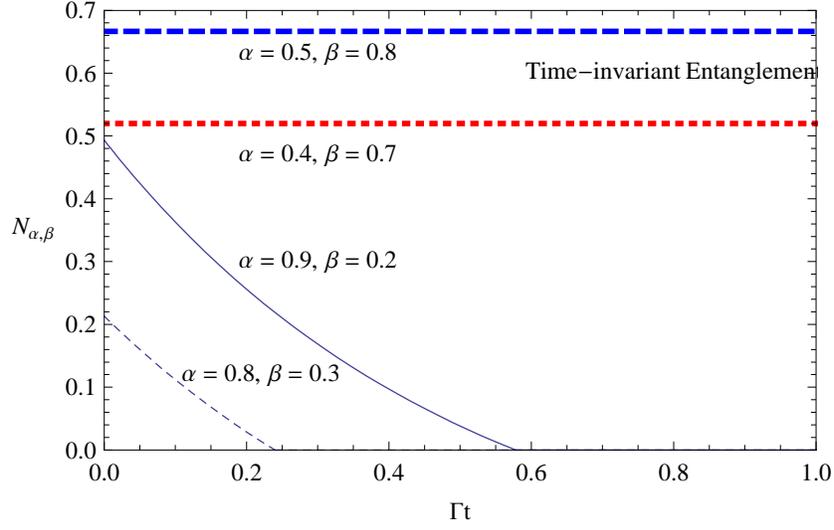}}
\caption{Negativity is plotted against parameter $\Gamma t$ for different sets of $\alpha$ and $\beta$. We observe time-invariant entanglement as well 
as finite time end of entanglement depending on range of these two parameters.}
\label{Fig:5}
\end{figure}
In Figure (\ref{Fig:5}), we plot negativity against parameter $\Gamma t$ for four different set of values of $\alpha$ and $\beta$. We see that 
for $\beta > 1/2$, that is, for $\alpha = 0.4$, $\beta = 0.7$ (red thick dashed line) and $\alpha = 0.5$, $\beta = 0.8$ (blue thick dashed line),  
we get time-invariant entanglement on the one hand and for other range, $\alpha = 0.9$, $\beta = 0.2$ (solid line), and 
$\alpha = 0.8$, $\beta = 0.3$ (thin dashed line), we see end of negativity at finite times. 

\begin{figure}[t!]
\scalebox{2.20}{\includegraphics[width=1.95in]{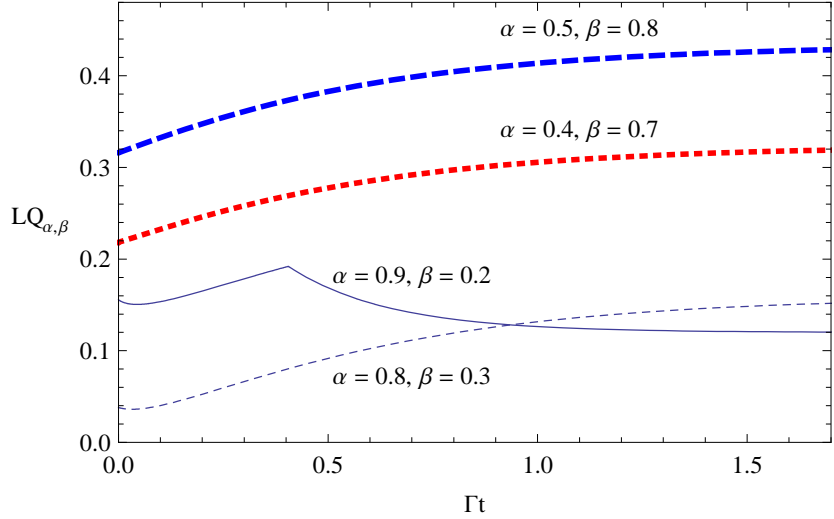}}
\caption{Local quantum uncertainity is plotted against parameter $\Gamma t$ for four set of values of $\alpha$ and $\beta$. It can be 
seen that in all cases, local quantum uncertainity tends to become stationary after exhibiting interesting dynamics at the start.}
\label{Fig:6}
\end{figure}

We have also calculated local quantum uncertainity for $\rho_{\alpha, \, \beta}(t)$. Following the procedure mentioned in previous section, we get 
a diagonal matrix and hence the eigenvalues of the resulting $3 \times 3$ matrix. It is simple to pick the maximum eigenvalue for given set of parameters. 
In Figure~(\ref{Fig:6}), we plot $LQ_{\alpha, \, \beta}$ against parameter $\Gamma t$ for same set of values for $\alpha$ and $\beta$ as in 
Figure~(\ref{Fig:5}). We observe quite interesting dynamics for local quantum uncertainity as compared with earlier cases. 
First we see that for two instances where we get time-invariant entanglement, the local quantum uncertainity first increases and then tends to 
freeze to a specfic positive value. Intuitively one can understand the freezing behavior of local quantum uncertainity as due to 
stationary correlations (not decaying due to decoherence subspace) in state $|\psi_3 \rangle$. However, it is not intuitive why these correlations first 
increase before becoming stationary. For the other two instances, where we get sudden death of entanglement, that is, for 
$(\alpha = 0.8, \, \beta = 0.3)$ (thin dashed line) local quantum uncertainity first decays for a short time and then once again increase and then tends 
to freeze to a constant value. Whereas for $(\alpha = 0.9, \, \beta = 0.2)$ (solid line), local quantum uncertainity first decreases for a short time, 
then increase to a value and then once again decays to another value and then finally exhibits freezing dynamics. As we mentioned, the freezing part of 
correlations can be explain easily whereas other parts of dynamics are counterintuitive. 

\subsubsection{Comparison with dynamics of random states}

In order to compare dynamics of quantum states with generic states, we have generated $100$ random pure states. 
A state vector for qubit-qutrit systems, randomly distributed according to the Haar measure can be generated in the following 
way \cite{Toth-CPC-2008}: First, we generate a vector such that both the real and the imaginary parts of the vector 
elements are Gaussian distributed random numbers with a zero mean and unit variance. Second we normalize the vector. It is easy to prove that the random 
vectors obtained this way are equally distributed on the unit sphere 
\cite{Toth-CPC-2008}. Note that the random pure states, which we generate in the global Hilbert space of dimension $6$, so the unit sphere is 
not the Bloch ball. 

After generation of $100$ random pure states, we find their time-evolved density matrices interacting with collective dephasing and 
compute negativity using PPT-mixture package \cite{Bastian-PRL106-2011}, for each state against parameter $\Gamma t$. From this data we can also obtain an 
error estimate to indicate the reliability of the measure. This can, for instance, be defined as a confidence interval \cite{Ali-JPB-2014}
\begin{eqnarray}
CI = \mu \, \pm \, \sqrt{\delta} \, ,
\label{interval}
\end{eqnarray}
where $\mu$ stands for mean value and $\delta$ for variance of quantity being measured. Note, however, that this is not a confidence
interval in the mathematical sense.

\begin{figure}[t!]
\scalebox{2.20}{\includegraphics[width=1.95in]{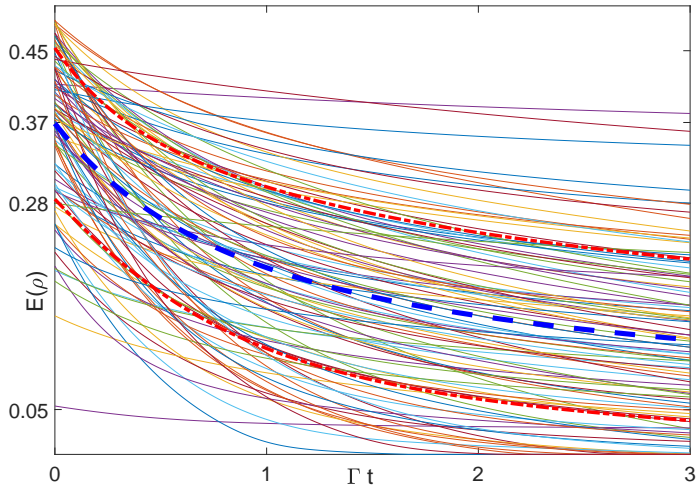}}
\caption{Entanglement monotone (negativity) is plotted against parameter $\Gamma t$ for $100$ initial random pure states. It can be 
seen that most of states remain NPT and approach to a fixed (freezing) value of entanglement after sufficiently long time.}
\label{Fig:7}
\end{figure}

In Figure (\ref{Fig:7}), we plot entanglement monotone (negativity) $E(\rho)$ for random pure states against parameter $\Gamma t$. The thick 
dashed (blue) line presents the mean value of entanglement, whereas thick dashed-dotted (red) lines represent confidence interval $CI$ with top line 
as sum of mean value and variance, where as below thick dashed-dotted line with difference of mean value and variance. As we can see that many states 
tend to exhibit freezing dynamics of entanglement (about $57\%$) where as many exhibit sudden death of entanglement (about $43\%$). 

\begin{figure}[t!]
\scalebox{2.20}{\includegraphics[width=1.95in]{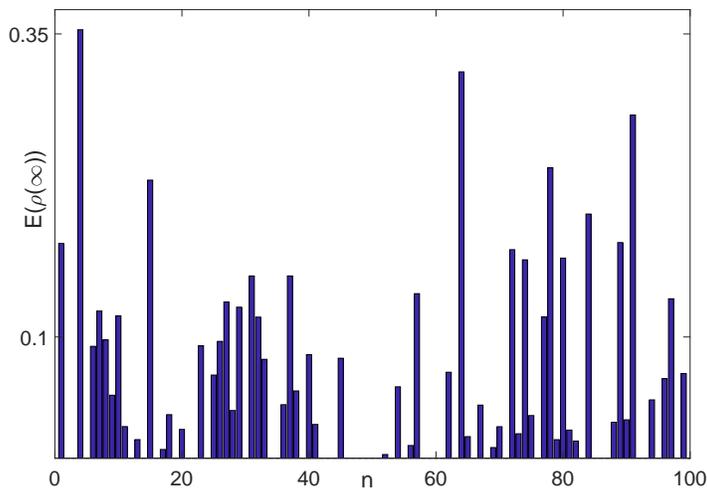}}
\caption{Entanglement monotone (negativity) is shown against number $n$ for $100$ initial random pure states. It can be 
seen that more than half of all states remain NPT.}
\label{Fig:8}
\end{figure}

Finally we analyze the asymptotic states by taking $\xi = 0$ in time-evolved density matrices for random states. We then compute entanglement monotone 
(negativity) for these states and as mentioned earlier about $57\%$ of them are found to be entangled. In Figure~(\ref{Fig:8}) we show bar graph for 
random states at infinity against number of random states. It is obvious that all entangled states will be having nonzero local quantum uncertainity 
as well.  

\section{Conclusions}
\label{Sec: Cc}

We have studied dynamics of quantum correlations of qubit-qutrit systems under Markovian collective dephasing. 
We have investigated some aspects of this simple system not studied before. In particular, we have studied two non-trivial features of 
entanglement dynamics, namely, time-invariant entanglement and freezing dynamics of entanglement. All previous studies on these two features of 
entanglement dynamics for bipartite as well as for multipartite quantum systems gave the impression that we could not have both features available 
for one specific quantum system under collective dephasing. The reason for this impression was the observation that for qubit-qubit systems 
we detected time-invariant entanglement whereas we did not find any freezing dynamics of entanglement under same collective dephasing model \cite{Liu-arXiv}. 
We did find freezing dynamics for qubit-qubit systems however for more general directions of magnetic fields \cite{Carnio-PRL-2015} instead of 
specific z-direction where we have only time-invariant feature available. For three qubits, we found evidence for freezing dynamics of genuine 
entanglement whereas we found no evidence for time-invariant entanglement \cite{Ali-EPJD-2017}. On the other hand, for four qubits, we found no 
evidence for freezing dynamics of entanglement but we do found time-invariant entanglement \cite{Ali-EPJD-2017}. 
More recently, we examined qutrit-qutrit quantum systems where we found freezing dynamics of entanglement but no time-invariant 
entanglement \cite{Ali-MPLA-2019}. There is no concrete mathematical arguments for mutual exclusiveness of these features for any specific 
Hilbert space. Contrary to earlier impression, for qubit-qutrit quantum systems, we found time-invariant entanglement as well 
as freezing dynamics entanglement. The future investigations might shed more light on relationship between these possibilities and dimensions of subsystems 
if there is any such relationship. 
In addition, we have studied dynamics of quantum discord for a specific family of quantum states and local quantum uncertainity for several families of 
states. We have seen that for some states quantum discord and local quantum uncertainity decay asymptotically and become zero only at infinity. For these 
states only classical correlations remain constant and do not decay. For other states which exhibit freezing dynamics of entanglement, local 
quantum uncertainity also tends to exhibit freezing dynamics. For quantum states which exhibit time-invariant entanglement, local quantum uncertainity first 
increase to a specific value and then become stationary at nonzero values. For same states which exhibit sudden death of entanglement, local quantum 
uncertainity first decay for short time, then increase for some time and finally reach a nonzero stationary value. 
Finally we have compared the dynamics of specific states with generic states by generating random pure states. We have seen that most of random pure states 
under collective dephasing exhibit freezing dynamics of entanglement and maintain this nonzero value even at infinity. Some random pure states do become 
separable at finite time. Another future avenue would be to explore more general $d \otimes N$ quantum systems for $d \neq N$ to find more examples.  




\end{document}